\documentclass[a4paper]{jpconf}
\usepackage{graphicx}
\begin{document}
\vspace*{-15mm}
\begin{flushright}
{CERN-PH-TH/2007-193}\\
\end{flushright}

\title{Enhanced Higgs boson production 
and avoidance of CP-violation and FCNC in the MPP inspired 2HDM
}

\author{C D Froggatt$^1$, R Nevzorov$^1$ and H B Nielsen$^2$}

\address{$^1$ Department of Physics and Astronomy, Glasgow University, Glasgow G12~8QQ, Scotland}
\address{$^2$ The Niels Bohr Institute, Copenhagen DK~2100, Denmark}

\ead{r.nevzorov@physics.gla.ac.uk}

\begin{abstract}
The multiple point principle (MPP) can be used to suppress non--diagonal flavour transitions
and CP violation in the two Higgs doublet extension of the standard model.
We discuss the quasi--fixed point scenario in the MPP inspired two Higgs doublet model
which leads to the enhanced production of Higgs particles at the LHC if the MPP scale 
$\Lambda\simeq 10-100\,\mbox{TeV}$.
\end{abstract}

\section{Multiple point principle and custodial symmetries}

The violation of CP invariance and the existence of tree--level flavor--changing neutral currents (FCNC) 
are generic features of $SU(2)_W\times U(1)_Y$ theories with two Higgs doublets. 
Indeed the couplings $m_3^2$, $\lambda_5$, $\lambda_6$ and $\lambda_7$ in the Higgs potential 
of the two--Higgs doublet model (2HDM) 
\begin{equation}
\begin{array}{c}
V= m_1^2 |H_1|^2 + m_2^2 |H_2|^2 -\biggl[m_3^2 H_1^{\dagger}H_2+h.c.\biggr]+
\frac{\lambda_1}{2} |H_1|^4 + \frac{\lambda_2}{2}|H_2|^4+\lambda_3 |H_1|^2 |H_2|^2\\ 
+\lambda_4|H_1^{\dagger}H_2|^2 + \biggl[\frac{\lambda_5}{2}(H_1^{\dagger}H_2)^2 + 
\lambda_6 |H_1|^2 (H_1^{\dagger}H_2)+\lambda_7 |H_2|^2(H_1^{\dagger}H_2)+h.c. \biggr]
\end{array}
\label{1}
\end{equation}
can be complex inducing CP--violation. Although one can eliminate the violation of CP invariance 
and tree--level FCNC transitions by imposing a discrete $Z_2$ symmetry, such a symmetry leads to the 
formation of domain walls in the early Universe which would create unacceptably large anisotropies in 
the cosmic microwave background radiation.

Here we consider the multiple point principle (MPP) \cite{1} as a possible mechanism for the 
suppression of FCNC and CP--violation effects. MPP postulates the existence of the maximal number 
of phases with the same energy density which are allowed by a given theory \cite{1}. When applied 
to the 2HDM the multiple point principle implies the existence of a large set of degenerate vacua at some 
high energy scale $\Lambda$ (MPP scale) that can be parametrised as
\begin{equation}
<H_1>=\left(
\begin{array}{c}
0\\ \Phi_1
\end{array}
\right)\; , \qquad <H_2>=\left(
\begin{array}{c}
0\\ \Phi_2\, e^{i\omega}
\end{array}
\right)\,,
\label{2}
\end{equation}
where $\Phi_1=\Lambda\cos\gamma$, $\Phi_2=\Lambda\sin\gamma$, $\tan\gamma=\lambda_1/\lambda_2$ while
$\omega$ is an arbitrary parameter. According to the MPP vacua at the electroweak and MPP scales must 
have the same vacuum energy density. 

The degeneracy of the MPP scale vacua can be achieved only if the Lagrangian for the Higgs--fermion 
interactions is invariant under symmetry transformations (see \cite{2}):
\begin{equation}
\begin{array}{c}
H_1\to e^{i\alpha}\,H_1, \qquad\quad u'_{R_i}\to e^{i\alpha}\,u'_{R_i}, \qquad\quad
d'_{R_i}\to e^{-i\alpha}\,d'_{R_i},\qquad\quad e'_{R_i}\to e^{-i\alpha}\,e'_{R_i},\\
H_2\to e^{i\beta}\,H_2, \qquad\quad u''_{R_i}\to e^{i\beta}\,u''_{R_i}, \qquad\quad
d''_{R_i}\to e^{-i\beta}\,d''_{R_i},\qquad\quad e''_{R_i}\to e^{-i\beta}\,e''_{R_i},
\end{array}
\label{3}
\end{equation}
where $u'_{i}$, $d'_{i}$, $e'_{i}$ are quark and lepton eigenstates which couple to $H_1$ while 
$u''_{i}$, $d''_{i}$, $e''_{i}$ are fermion eigenstates that interact with $H_2$. In Eq.~(\ref{3}) the 
subscript $R$ denote right-handed fermion fields. Two global $U(1)$ symmetries (\ref{3}) forbid non--diagonal 
flavour transitions at the tree level. Moreover these symmetries lead to the vanishing of $\lambda_5$, 
$\lambda_6$ and $\lambda_7$ in the Higgs potential (\ref{1}) that cause CP--nonconservation in the 2HDM. 
The mixing term $m_3^2(H_1^{\dagger}H_2)$, which is not forbidden by the MPP, softly breaks custodial 
global symmetries but does not create new sources of CP--violation or FCNC transitions.

The degeneracy of the physical and MPP scale vacua requires Higgs and Yukawa couplings to be adjusted 
so that an appropriate cancellation among the quartic terms in the Higgs potential (\ref{1}) takes place. 
Such cancellation becomes possible only if $\lambda_4(\Lambda)<0$ and the combination of the Higgs 
self--couplings $\tilde{\lambda}=\sqrt{\lambda_1 \lambda_2}+\lambda_3+\lambda_4$ as well as its beta 
function $\beta_{\tilde{\lambda}}$ go to zero at the scale $\Lambda$ \cite{3}. At moderate values of 
$\tan\beta=v_2/v_1$ where $v_2$ and $v_1$ are vacuum expectation values of $H_2$ and $H_1$ in the 
physical vacuum, the last two MPP conditions can be written as follows \cite{3} 
\begin{equation}
\begin{array}{c}
\lambda_3(\Lambda)=-\sqrt{\lambda_1(\Lambda)\lambda_2(\Lambda)}-\lambda_4(\Lambda)\,,\\
\lambda_4^2(\Lambda)=\displaystyle\frac{6h_t^4(\Lambda)\lambda_1(\Lambda)}{(\sqrt{\lambda_1(\Lambda)}+\sqrt{\lambda_2(\Lambda)})^2}
-2\lambda_1(\Lambda)\lambda_2(\Lambda)-\frac{3}{8}\biggl(3g_2^4(\Lambda)+2g_2^2(\Lambda)g_1^2(\Lambda)+g_1^4(\Lambda)\biggr)\,,
\end{array}
\label{4}
\end{equation}
where $h_t(\Lambda)$ is the top quark Yukawa coupling. Thus $\lambda_3$ and $\lambda_4$ are not independent parameters 
in the considered model. As a result the MPP inspired 2HDM has less free parameters than the 2HDM of type II and 
therefore can be regarded as a minimal non--supersymmetric two Higgs doublet extension of the SM. 
It is also worth noting here that the MPP conditions (\ref{4}) are satisfied identically in the MSSM
above the supersymmetry breaking scale.

The MPP conditions (\ref{4}) must be supplemented by the vacuum stability requirements, i.e. $\lambda_{1,\,2}(\mu)>0$ and 
$\tilde{\lambda}(\mu)>0$, which have to be fulfilled everywhere from $\mu=M_t$ to $\mu=\Lambda$.
Otherwise another minimum of the Higgs potential (\ref{1}) with negative vacuum energy density arises at some 
intermediate scale, preventing the consistent realisation of the MPP in the 2HDM.

\section{Higgs phenomenology}

The renormalisation group (RG) flow of all couplings in the MPP inspired 2HDM is determined by $\lambda_1(\Lambda)$, 
$\lambda_2(\Lambda)$ and $h_t(\Lambda)$. When $h_t(\Lambda)> 1$ the solutions of the RG equations for the top quark 
Yukawa coupling are concentrated in the vicinity of the quasi--fixed point at the electroweak scale. The value
of $\tan\beta$ that corresponds to the quasi--fixed point scenario depends mainly on the MPP scale $\Lambda$.
It varies from $1.1$ to $0.5$ when $\Lambda$ changes from $M_{Pl}$ to $10\,\mbox{TeV}$ \cite{4}. At large values of $h_t(\Lambda)$ 
the MPP and vacuum stability conditions constrain $\lambda_i(\Lambda)$ very strongly. 
Our numerical studies show that for $\Lambda=M_{Pl}$ and 
$\lambda_1(M_{Pl})=\lambda_2(M_{Pl})=\lambda_0$ the ratio $\lambda_0/h_t^2(M_{Pl})$ can vary only within a very narrow 
interval from $0.79$ to $0.87$ if $h_t(\Lambda)> 1.5$. This ensures the convergence of the solutions of the RG equations for 
$\lambda_i(\mu)$ to the quasi--fixed points.

\begin{figure}[h]
\hspace*{-0cm}{$m_{h_i,\,\chi^{\pm}}$}\hspace*{7.2cm}{$|R_{t\bar{t}h_i}|$}\\
\begin{minipage}{18pc}
\includegraphics[width=18pc]{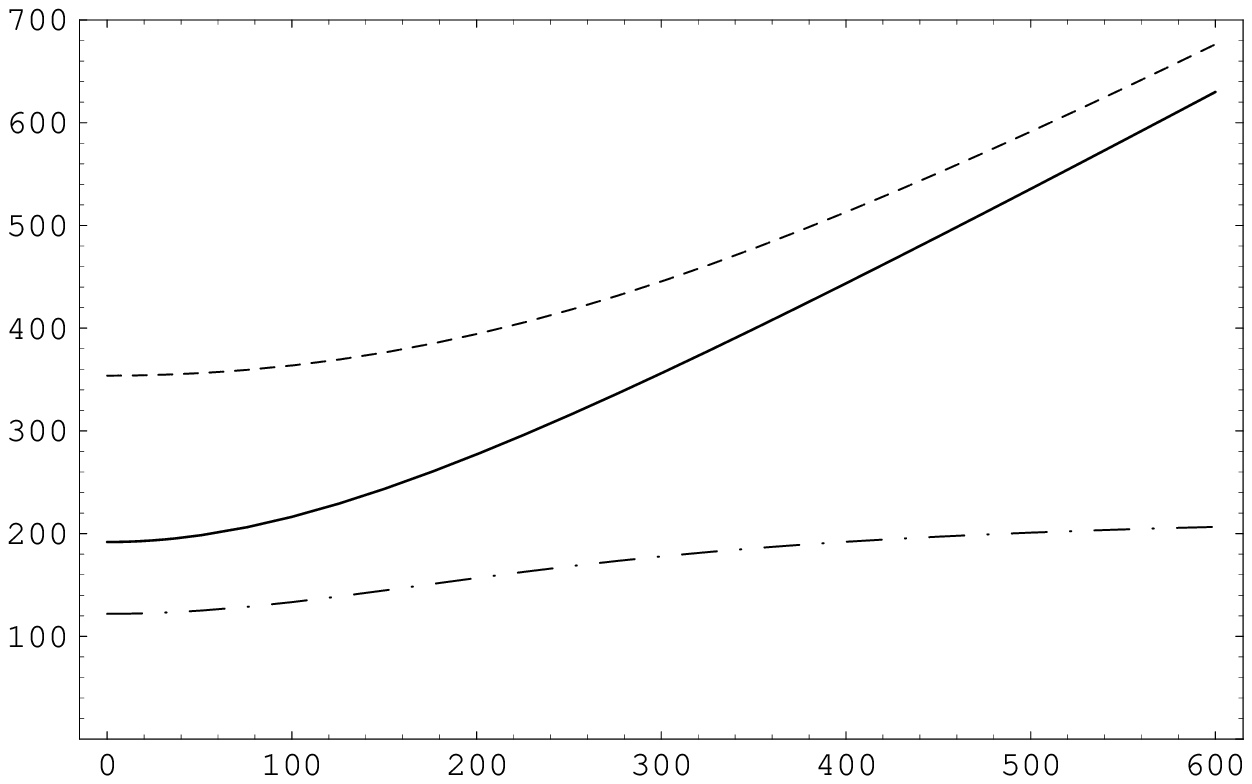}\\
\hspace*{4cm}{$m_A$}
\caption{\label{label} Spectrum of Higgs bosons near the quasi--fixed point versus $m_A$ for $\Lambda=10\,\mbox{TeV}$. 
The dash--dotted and dashed lines correspond to the CP--even Higgs boson masses while the solid line represents 
the mass of the charged Higgs states.
}
\end{minipage}\hspace{2pc}%
\begin{minipage}{18pc}
\includegraphics[width=18pc]{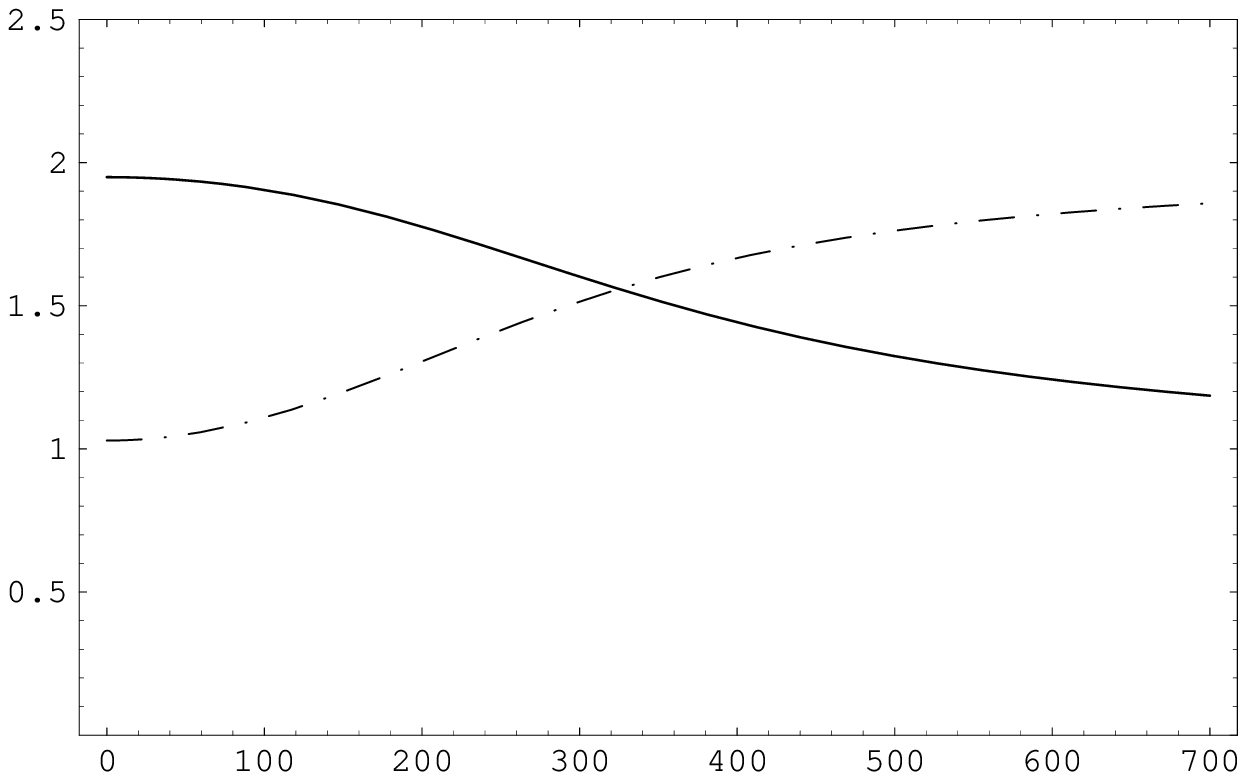}\\
\hspace*{4cm}{$m_A$}
\caption{\label{label} Absolute values of the relative couplings $R_{t\bar{t}i}$ of the Higgs scalars to
the top quark in the quasi--fixed point scenario as a function of $m_A$ for $\Lambda=10\,\mbox{TeV}$. Solid 
and dashed--dotted curves correspond to the lightest and heaviest CP--even Higgs states.
}
\end{minipage}\\ 
\end{figure}

The Higgs spectrum of the two Higgs doublet extension of the SM contains two charged and three neutral scalar states. 
Because in the MPP inspired 2HDM CP--invariance is preserved one of the neutral Higgs bosons is always CP--odd while 
the two others are CP--even. The qualitative pattern of the Higgs spectrum depends very strongly on the mass of the 
pseudoscalar Higgs boson $m_A$ (see Fig.~1). When $m_A\gg M_t$ the masses of charged, CP--odd and heaviest CP--even
Higgs bosons are almost degenerate around $m_A$. In the considered limit the lightest CP--even Higgs boson
mass $m_{h_1}$ attains its maximal value which is determined by $\lambda_i(M_t)$ and $\tan\beta$. 

For each MPP scale $\Lambda$ one can compute the values of the Higgs self--couplings at the electroweak scale 
and $\tan\beta$ near the quasi--fixed point. Because at large values of the pseudoscalar mass $m_{h_1}$ is almost independent 
of $m_A$, the upper bound on the lightest Higgs scalar mass depends predominantly on the scale $\Lambda$. When the MPP scale 
is high such dependence is relatively weak. If $\Lambda> 10^{10}\,\mbox{GeV}$ the mass of the lightest Higgs particle 
in the quasi--fixed point scenario does not exceed $125\,\mbox{GeV}$ \cite{4}. 
However at low MPP scales $\Lambda\simeq 10-100\,\mbox{TeV}$ the theoretical restriction on $m_{h_1}$ reaches $200-220\,\mbox{GeV}$. 
The lightest Higgs scalar in the considered case is predominantly a SM--like Higgs boson, since its relative coupling 
to a $Z$ pair is rather close to unity. Nevertheless at low MPP scales the quasi--fixed point scenario leads to large 
values of the relative coupling of the lightest Higgs scalar to the top quark resulting in the enhanced production 
of this particle at hadron colliders (see Fig.~2). Thus the analysis of production and decay rates of the SM--like Higgs boson at 
the LHC should make possible the distinction between the quasi--fixed point scenario in the MPP inspired 2HDM with low 
scale $\Lambda$, the SM and the MSSM even if extra Higgs states are relatively heavy, i.e. $m_A\simeq 500-700\,\mbox{GeV}$. 

\section{Conclusions}
We have argued that the MPP assumption allows us to suppress FCNC and CP violating phenomena in the 2HDM.
When $h_t(\Lambda)>1$ the solutions of the RG equations in the MPP inspired 2HDM converge to the quasi--fixed point,
leading to stringent restrictions on the lightest Higgs scalar mass. In the quasi--fixed point scenario
the Higgs couplings to the $t$--quark can be significantly larger than in the SM, which allows us to test this scenario 
at hadron colliders.

\ack
RN acknowledge support from the SHEFC grant HR03020 SUPA 36878.

\end{document}